\documentclass[aps,prl,twocolumn,superscriptaddress,showpacs,amsmath,amssymb,preprintnumbers]{revtex4}

\usepackage{graphicx,amssymb}

\begin{document}

\preprint{Ver 3.2x}

\title{Huge Transverse Magnetization in the Field-Induced Phase of the Antiferromagnetic Molecular Wheel CsFe$_8$}

\author{L. Schnelzer}
\affiliation{Physikalisches Institut, Universit\"{a}t Karlsruhe (TH), 76128 Karlsruhe, Germany}

\author{O. Waldmann}
\affiliation{Department of Chemistry and Biochemistry, University of Bern, 3012 Bern, Switzerland}

\author{M. Horvati\'{c}}
\affiliation{Grenoble High Magnetic Field Laboratory, CNRS, BP 166, 38042 Grenoble Cedex 9, France}

\author{S. T. Ochsenbein}
\affiliation{Department of Chemistry and Biochemistry, University of Bern, 3012 Bern, Switzerland}

\author{S. Kr\"{a}mer}
\affiliation{Grenoble High Magnetic Field Laboratory, CNRS, BP 166, 38042 Grenoble Cedex 9, France}

\author{C. Berthier}
\affiliation{Grenoble High Magnetic Field Laboratory, CNRS, BP 166, 38042 Grenoble Cedex 9, France}
\author{H. U. G\"{u}del}
\affiliation{Department of Chemistry and Biochemistry, University of Bern, 3012 Bern, Switzerland}

\author{B. Pilawa}
\affiliation{Physikalisches Institut, Universit\"{a}t Karlsruhe (TH), 76128 Karlsruhe, Germany}

\date{\today}

\begin{abstract}
The $^1$H-NMR spectrum and nuclear relaxation rate $T_1^{-1}$ in the antiferromagnetic wheel CsFe$_8$ were
measured to characterize the previously observed magnetic field-induced low-temperature phase around the
level crossing at 8~T. The data show that the phase is characterized by a huge staggered transverse
polarization of the electronic Fe spins, and the opening of a gap, providing microscopic evidence for the
interpretation of the phase as a field-induced magneto-elastic instability.
\end{abstract}


\maketitle

\begin{figure}[hb]
\includegraphics{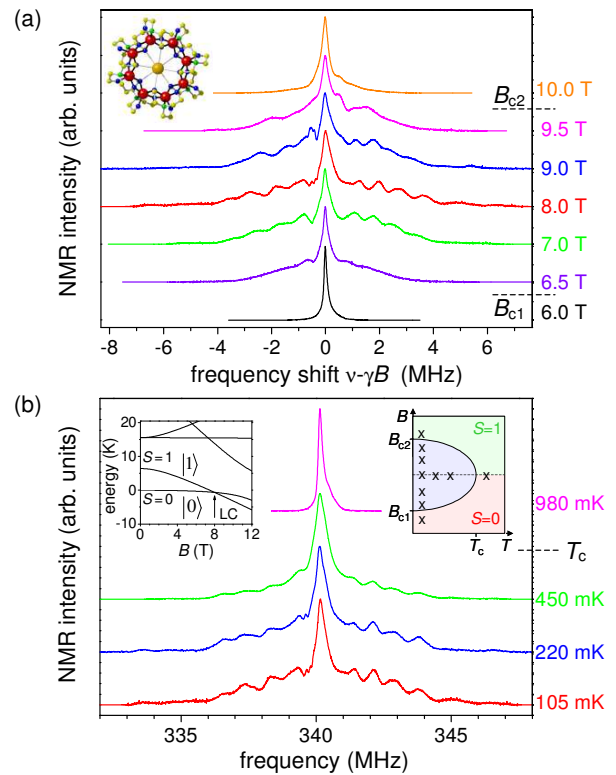}
\caption{\label{fig:one} (color online) (a) Field dependence at $T = 105$~mK and (b) temperature dependence
at $B = 8$~T of the $^1$H-NMR spectrum of CsFe$_8$. The inset in (a) shows the structure of CsFe$_8$, the
left inset in (b) the energy spectrum, and the right inset in (b) the phase diagram near the level crossing;
crosses indicate the $(B,T)$ values for which spectra are shown.}
\end{figure}

Molecular magnetic clusters of nanometer size have received enormous attention recently because of their
spectacular quantum phenomena \cite{Mn12_Fe8}. A peculiar class of magnetic cluster is the antiferromagnetic
(AFM) molecular wheel, in which magnetic metal ions are assembled in a ring-like structure [inset to
Fig.~\ref{fig:one}(a)]\cite{Wheels}. Strong AFM Heisenberg interactions between the metal ions lead to a
nonmagnetic $S = 0$ ground state and a first excited $S = 1$ state in zero magnetic field. In a field the
Zeeman splitting lifts the degeneracy of the $S=1$ level, such that a sufficiently strong field induces a
level crossing (LC), where the ground state of the molecule changes from $S = 0$, $M = 0$ to $S = 1$, $M =
-1$ [inset to Fig.~\ref{fig:one}(b)] \cite{Wheels,Cornia99,Waldmann9901}. Due to the quasi degeneracy near
the LC, the magnetism of the molecule is sensitive to weak interactions with the environment, such as the
lattice vibrations or the surrounding magnetic molecules. Studying the LCs in AFM wheels is hence of broad
interest as it provides experimental insight into the interplay of a mesoscopic quantum system with its
environment.

The molecule [CsFe$_8$\{N(CH$_2$CH$_2$O)$_3$\}$_8$]Cl, or CsFe$_8$ in short \cite{Saal97}, which realizes a
ring of eight spin-5/2 Fe(III) ions, is particularly interesting as it is the only AFM wheel system in which
magnetic torque studies revealed a phase transition below a temperature $T_c$ of about 0.7~K and in a field
range of about $\pm 1.5$~T around the LC, between fields $B_{c1}$ and $B_{c2}$ [the phase diagram is sketched
in the inset to Fig.~\ref{fig:one}(b)] \cite{Waldmann06,PT}. This could be a field-induced magnetic phase
transition, similar to the weakly-interacting dimer compounds, such as TlCuCl$_3$, where a Bose-Einstein
condensation of magnons occurs \cite{Nikuni00}. However, the prerequisite magnetic interactions between
clusters are negligible in CsFe$_8$, as evidenced by the crystal structure and other arguments
\cite{Waldmann06}. The novel scenario of a field-induced magneto-elastic instability, or spin 
Jahn-Teller effect was hence proposed \cite{Waldmann06,Waldmann07}, which explained the torque data well.

In order to obtain microscopic insight into this phase, we performed a proton nuclear magnetic resonance
(NMR) study of CsFe$_8$ single crystals. The $^1$H-NMR spectrum and $T_1^{-1}$ relaxation rate allow one to
investigate the local spin configuration and spin dynamics, respectively, of the Fe spins in a wheel
\cite{Julien99,Pilawa05}. The data provide the first strong experimental evidence for the proposed
field-induced magneto-elastic instability in CsFe$_8$.

Crystals of CsFe$_8$$\cdot$5CHCl$_3$$\cdot$0.5H$_2$O were synthesized following \cite{Saal97}. The space
group is $P2_1/n$; the molecules exhibit approximate $C_4$ symmetry along the wheel or crystal $c$ axis. The
NMR measurements were performed at the Grenoble High Magnetic Field Laboratory. The single-crystal samples
were mounted in the mixing chamber of a dilution refrigerator to ensure efficient thermalization. The
$^1$H-NMR spectra were measured by a $\frac{\pi}{2}$-$\tau$-$\pi$ echo sequence, with $\tau$ times as short
as 3~$\mu$s, much shorter than the $T_2$ relaxation time, to minimize $T_2$ effects \cite{Abragam61}. The
$T_1^{-1}$ relaxation rates were measured by an Inversion-Recovery (three-pulse) sequence \cite{Abragam61}.
In total six samples were measured. The results presented here were recorded on a crystal with an angle
between $c$ axis and magnetic field of $\varphi = 95(2)^\circ$. From NMR measurements not shown here the
critical values were determined to $B_{c1} = 6.4$~T, $B_{c2} = 9.7$~T, and $T_c = 0.75$~K.

\begin{figure}
\includegraphics{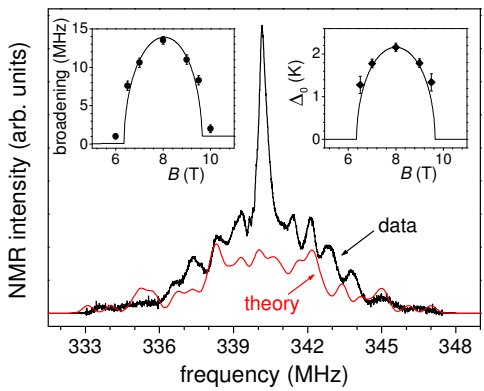}
\caption{ \label{fig:two} (color online) $^1$H-NMR spectrum at 105~mK and 8~T and the simulation. The left and right insets
show the broadening of the spectrum and the deduced gap $\Delta_0$ (symbols), respectively, as well as the
theoretical expectation (lines).}
\end{figure}

Figure~\ref{fig:one}(a) presents the $^1$H-NMR spectra at 105~mK for fields from 6 to 10~T. Because of the
gyromagnetic ratio $\gamma = 42.537$~MHz/T the central line shifts with magnetic field; the spectra are hence
displayed relative to $\nu_0 = \gamma B$. At fields below the lower critical field $B_{c1}$, a single NMR
line is observed with a full width at half maximum (FWHM) of 0.5~MHz. Above the upper critical field
$B_{c2}$, the spectrum consists of a somewhat broadened central line with a small shoulder. This is the
expected behavior for AFM wheels, as observed also before in other wheels \cite{Pilawa05}: for small fields
CsFe$_8$ is in the $S = 0$ state and the line width reflects the dipolar couplings between the protons. For
large fields the $S = 1$ state is stabilized and the additional couplings between the protons and Fe spins
result in a broadening of the line. For fields between $B_{c1}$ and $B_{c2}$, the spectra are characterized
by a huge broadening of up to 14~MHz: several pronounced satellite lines appear in the frequency range of
$\pm 5$~MHz around the main peak, with weaker lines extending up to $\pm 7$~MHz (Fig.~\ref{fig:two}).
Figure~\ref{fig:one}(b) shows the temperature dependence: the structure in the spectra remains essentially
unchanged and disappears above $T_c$.

In the following we focus on the unprecedentedly large broadening of the $^1$H-NMR spectrum in the regime of
the field-induced phase. It will be shown that it is a direct signature of a huge transverse magnetic
polarization of the CsFe$_8$ molecule. Since the broadening is symmetric, the transverse polarization has to
be staggered, consistent with AFM interactions in the wheel.

The dipolar coupling of the 96 nuclear $^1$H spins and the 8 electronic Fe spins in a CsFe$_8$ molecule is
described by the Hamiltonian \cite{Abragam61}
\begin{eqnarray}
 \label{eq:one}
 \hat{\cal{H}}^{NMR} &=&
  -\sum_k \hbar \gamma {\bf B} \cdot \hat{\bf I}_k
  + \sum_{ki} \hat{\bf I}_k \cdot {\bf D}_{ki} \cdot \hat{\bf S}_i,
  \\
  D^{\nu \mu}_{ki} &=&
  -\frac{ 78.973\text{ MHz\AA}^3 }{R^3_{ki}} ( \delta_{\nu \mu} - 3 \frac{ R_{ki,\nu} R_{ki,\mu} }{ R^2_{ki} }
  ),
\end{eqnarray}
where $k$ refers to the protons and $i$ to the Fe ions, and $\nu,\mu = x,y,z$. The first, dominant term is
the nuclear Zeeman energy. The second term represents the dipolar coupling between the $^1$H nuclear spins
$\hat{\bf I}_k$ and Fe spins $\hat{\bf S}_i$, where ${\bf R}_{ki}$ denotes the distance vector between the
two spins, which is known from the crystal structure. The chemical shift at each proton site can be omitted
as it is completely negligible compared to the huge spectral broadening in the field-induced phase. Also, the
\textit{a priori} unknown transferred hyperfine coupling has been neglected; we will come back to this later
in the discussion. In a strong magnetic field, and at zero temperature, the resonance frequency of the $k$th
proton is shifted with respect to $\nu_0$ by
\begin{eqnarray}
 \label{eq:two}
 \delta \nu_k =
  - \sum_{i} D^{zz}_{ki} \langle B| \hat{S}_{iz} |B\rangle  - \sum_{i} D^{zx}_{ki} \langle B| \hat{S}_{ix} |B\rangle.
\end{eqnarray}
Here, $z$ denotes the magnetic field direction, and $|B\rangle$ the (field-dependent) ground state of
CsFe$_8$. The $x$ and $y$ directions were chosen such that $\langle B|\hat{S}_{iy}|B\rangle = 0$. The first
term in Eq.~(\ref{eq:two}) reflects the shift due to a longitudinal magnetic polarization, and the second one
that due to a transverse polarization. It is noted that the spin polarization detected by NMR is static on
the time scale of the NMR experiment \cite{Abragam61}.

At low fields, CsFe$_8$ is non-magnetic and the shifts $\delta \nu_k$ are zero. At high fields, the molecule
is in the $S = 1$ state, which is characterized by a uniform longitudinal polarization of size 1/8~$g\mu_B$
on each Fe ion, i.e., $\langle B|\hat{S}_{iz}|B\rangle = -1/8$ and $\langle B|\hat{S}_{ix}|B\rangle = 0$. The
smallest distance between a proton and an Fe ion in CsFe$_8$ is 3.05~{\AA}; the shift by the longitudinal
polarization is thus bounded above by 0.7~MHz, in perfect accordance with the experiment at high fields. The
bound, however, implies that the ten times broader NMR spectra observed in the field-induced phase cannot be
accounted for by a longitudinal polarization. A transverse polarization is hence considered. This component
can be indeed obtained by a mixing of the two states at the LC, which will be denoted as $|0\rangle$ and
$|1\rangle$, see left inset to Fig.~\ref{fig:one}(b). We write $|B\rangle = \alpha(B) |0\rangle + \beta(B)
|1\rangle$, with $\alpha^2+\beta^2=1$. At zero temperature the longitudinal polarization is then $\langle
B|\hat{S}_{iz}|B\rangle = \beta^2 \langle 1|\hat{S}_{iz}|1\rangle$. The mixing, however, also induces a
transverse polarization $\langle B|\hat{S}_{ix}|B\rangle = 2 \alpha\beta \langle 0|\hat{S}_{ix}|1\rangle$.

The matrix elements were evaluated as follows. First, for $\langle 1|\hat{S}_{iz}|1\rangle$ and $\langle
0|\hat{S}_{ix}|1\rangle$ one has to solve the (electronic) spin Hamiltonian of CsFe$_8$, $\hat{\cal{H}} = -J
\sum_i \hat{\bf S}_i \cdot \hat{\bf S}_{i+1} + D \sum_i \hat{S}_{ic}^2 + \mu_B g {\bf B} \cdot \hat{\bf S}$,
which consists of the AFM nearest-neighbor Heisenberg interactions in the wheel ($J = -21.5$~K), the
easy-axis single-ion anisotropy ($D = -0.55$~K), and the electronic Zeeman term ($\hat{\bf S} = \sum_i
\hat{\bf S}_i$) \cite{Waldmann9901,Waldmann06b}. This was achieved by numerical diagonalization of
$\hat{\cal{H}}$ in the two-sublattice approximation (which works excellently for the states relevant here)
\cite{Waldmann06b}, for given field $B$ and orientation $\varphi$. Then, the local spin matrix elements were
evaluated for given $\alpha,\beta$. In order to mimic the field-induced phase, the mixing was chosen such
that $\beta^2$ is zero for fields below $B_{c1}$, increases linearly with field from 0 to 1 in the range
$B_{c1}$ to $B_{c2}$, and is 1 above $B_{c2}$. This ensures that outside the range [$B_{c1}$,$B_{c2}$] the
mixing (and transverse polarization) is zero. The linear increase between $B_{c1}$ and $B_{c2}$ is suggested
by the experimentally observed linear increase of the magnetization in the field-induced phase, i.e., of
$-\langle B|\hat{S}_{z}|B\rangle = -\beta^2 \sum_i \langle 1|\hat{S}_{iz}|1\rangle$. At exactly the LC field,
$\beta^2 = 1/2$ and mixing is maximal.

\begin{figure}
\includegraphics{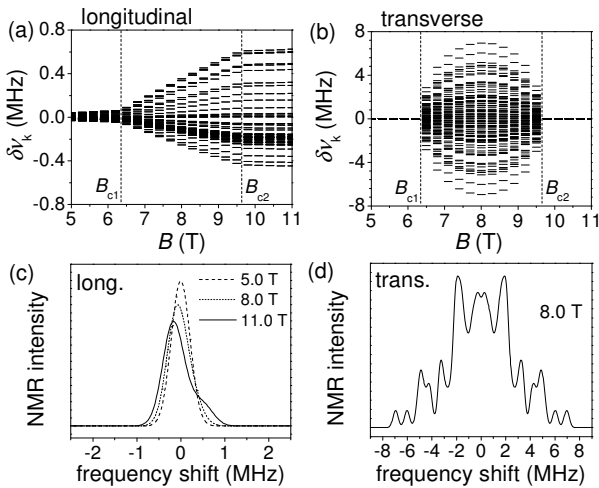}
\caption{\label{fig:three} Simulation of the NMR shifts $\delta \nu_k$ due to the (a) longitudinal and (b)
transverse polarization, and (c,d) the corresponding NMR spectra after convolution with a Gaussian (FWHM =
0.5~MHz) for selected fields. For comparison with experiment, the shifts in (a) and (b) have to be added.}
\end{figure}

The upper panels in Fig.~\ref{fig:three} present the calculated contributions of the longitudinal and
transverse polarizations to the NMR shifts $\delta \nu_k$. The curves in the lower panels of
Fig.~\ref{fig:three} were obtained from the data in the upper panels by convolution with a Gaussian to
account for the NMR line width. For the longitudinal polarization, the calculation yields asymmetric shifts
of at most 0.6~MHz, in accordance with the above bound. In the low-field regime, the spectrum consists of a
narrow peak. With increasing field the central line is slightly shifted and a weak shoulder develops, in very
good agreement with the experiments outside the field-induced phase. As regards the transverse polarization,
the calculation reveals very large shifts of up to 7~MHz, which build up quickly even for small mixing. For
comparison with experiment, the shifts produced by the longitudinal and transverse polarization have to be
added. The result at the LC field, after convolution with a Gaussian, is shown in Fig.~\ref{fig:two}.

The comparison of experiment and simulation in Fig.~\ref{fig:two} shows an excellent agreement, providing a
reliable proof of the spin configuration in the field-induced phase. It is noted that the central peak in the
experimental spectra is mostly due to the protons in the solvent (outside the ring molecules), which were
omitted in the simulation. We want to stress that the calculations did not involve any free parameter,
besides assuming $\alpha = \beta$, which is certainly reasonable at the LC. The calculated transverse
polarization is staggered with a magnitude of 1.7 per Fe center (in units of $g\mu_B$). This is a stunning 27
times larger than the longitudinal polarization of $\beta^2/8 = 1/16$. Roughly speaking, the transverse
component is built up by almost the complete spin (5/2) at each site, while the longitudinal component comes
from $S=1$ ($\times \beta^2$) distributed over 8 sites. For comparison, in spin-1/2 dimer systems such as
TlCuCl$_3$, in which also a $S = 0$ and $S = 1$ level mix, but because of very different physics
\cite{Nikuni00}, the transverse polarization is at most 0.35, a mere 1.4 times larger than the longitudinal
polarization of 0.25 \cite{Momoi00}. Interestingly, also for $D = 0$ a large transversal magnetization of 1.1
per Fe is calculated. Hence, the essential factor in the huge broadening of the NMR spectrum is not $D$, but
the mixing of the $|0\rangle$ and $|1\rangle$ levels. However, the magnetic anisotropy is the key element in
a field-induced magneto-elastic instability \cite{Waldmann06}, and the size of $D$ may be the crucial factor
for the existence of the field-induced phase in CsFe$_8$ in the first place.

Some discrepancies between the observed and simulated spectra can be noted in Fig.~\ref{fig:two}. One
possible reason could be a transferred hyperfine coupling, which was neglected in Eq.~(\ref{eq:one}). This
coupling is isotropic, and its transverse component hence zero. Therefore, it does not contribute to the
dominating transverse component of the calculated NMR spectrum [shown in Fig.~\ref{fig:three}(d)]. It will,
however, contribute to the much smaller longitudinal component [shown in Fig.~\ref{fig:three}(c)], and thus
may slightly modify the details of the final calculated spectrum. A second reason could be the (implicit)
assumption of a structurally symmetric wheel in the spin Hamiltonian $\hat{\cal{H}}$. Some weak distortion of
the real molecule, either due to crystallographic reasons and/or the proposed field-induced magneto-elastic
instability, will produce slightly different local spin polarizations.

\begin{figure}
\includegraphics{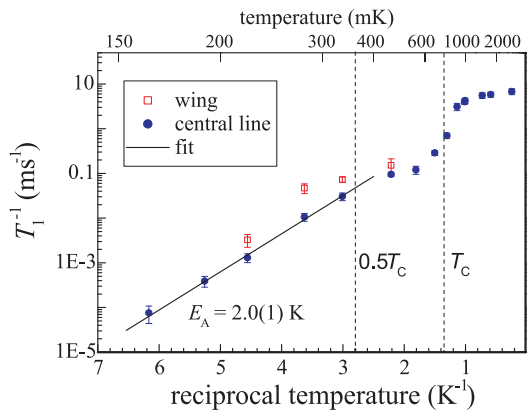}
\caption{\label{fig:four} (color online) $^1$H relaxation rate $T_1^{-1}$ vs. reciprocal temperature at 8~T
as measured on the central line and in the wing 3~MHz below the central line. The dashed lines indicate three
temperature regimes. The solid line corresponds to an Arrhenius fit of the low-$T$ data.}
\end{figure}

We now turn to the temperature dependence of the $^1$H nuclear $T_1^{-1}$ relaxation rate. The $T_1^{-1}$
rate is mainly driven by the fluctuations of the Fe spins \cite{Julien99,Pilawa05}, somewhat modified
(homogenized) by the nuclear spin diffusion mechanism \cite{Abragam61}. The $T_1^{-1}$ data measured at the
LC field are shown in Fig.~\ref{fig:four}. One set of data points was recorded at the central line at
340.15~MHz. Three different regimes can be identified. Above $T_c$, $T_1^{-1}$ is almost constant, with a
small decrease towards lower temperatures, as expected for any high-temperature regime. At $T_c$, the
relaxation rate drops significantly, indicating the transition to the field-induced phase. The drop is
enhanced by a strong reduction of the nuclear spin diffusion due to the huge broadening of the NMR spectrum.
Below about $0.5 T_c$, the $T_1^{-1}$ rate shows a thermally activated behavior, $T_1^{-1} \propto
\exp[-E_A/(k_B T)]$, with an activation energy of $E_A = 2.0(1)$~K.

In order to confirm that the $T_1^{-1}$ data at the central peak represent the field-induced phase well, some
control values were measured at 337.40~MHz, i.e., in the wing of the NMR spectrum, which exists only in the
field-induced phase (Fig.~\ref{fig:four}). The same temperature dependence was indeed observed. The rates are
somewhat enhanced, as expected for nuclei more strongly coupled to Fe spins.

Our NMR results provide a strong microscopic evidence for the previously proposed description of the
field-induced phase in CsFe$_8$ as a field-induced magneto-elastic instability \cite{Waldmann06}. The
thermodynamic torque data could be well reproduced by an effective two-level Hamiltonian, which includes the
important non-diagonal magneto-elastic couplings \cite{Waldmann06,Waldmann07}. The key prediction of this model is a
mixing of the two levels and a concomitant opening of an energy gap in the field-induced phase. The observed
huge broadening of the NMR spectrum was shown in the above to be directly related to a strong transverse
polarization due to a mixing of the two lowest spin states. These measurements hence provide a direct
microscopic confirmation of the level mixing in the field-induced phase. Furthermore, the model predicts a
thermally activated behavior at low temperatures, with an activation barrier equal to the energy gap. At the
LC, the model predicts a gap of $\Delta_0^{max} = g \mu_B (B_{c2}-B_{c1})/2$. For our sample this implies a
gap of 2.2~K, which is in excellent agreement with the activation energy $E_A = 2.0$~K deduced from the
$T_1^{-1}$ measurement.

Either one of the features: broadening, transverse polarization, or mixing may be considered as the order
parameter of the phase transition. Within a two-level Hamiltonian description, these features are related to
a gap parameter $\Delta_0 = 2 \langle 0| \hat{\cal{H}}|1\rangle$, which is most conveniently taken as the
order parameter. The field dependence of the broadening of the NMR spectra at 105~mK, and the $\Delta_0$
determined from it, are presented in the insets to Fig.~\ref{fig:two}. They show how the field-induced phase
develops as a function of field, in very good agreement with the model predictions. Furthermore, the model
provides a (rough) estimate of $T_c$, $k_B T_c = \Delta_0^{max}/2$, which is also consistent with experiment.
It is remarked that an energy gap at the LC of 2~K represents a rather strong effect on the magnetic degrees
of freedom in CsFe$_8$; this value should be compared to the gap of 6~K in zero field [inset to
Fig.~\ref{fig:one}(b)].

In summary, we studied the field and temperature dependence of the $^1$H-NMR spectrum and $T_1^{-1}$
relaxation rate in a single crystal of the molecular wheel CsFe$_8$ in the field-induced low-temperature
phase around the level crossing at 8.0~T. The NMR data unambiguously demonstrated that this phase is
characterized by (i) a huge broadening of the $^1$H-NMR spectra due to a huge staggered transverse
polarization of the Fe spins, and (ii) an energy gap of 2.0~K at the level crossing. These findings provide
strong microscopic evidence for the previously proposed model of a field-induced magneto-elastic instability.
Future work should attempt the direct measurement of structural distortions and determination of the relevant
vibration modes and spin-phonon coupling constants. Also, a microscopic theory is lacking.

\begin{acknowledgments}
Financial support by EC-RTN-QUEMOLNA, contract n$^\circ$ MRTN-CT-2003-504880, and the Swiss National Science
Foundation is acknowledged.
\end{acknowledgments}


\begin{references}
\bibitem{PT}
The magnetic torque data, as well as the NMR data, suggests that the observed field-induced phase is a phase
transition. However, according to very recent theoretical work it could also be a sharp crossover. The
presently available data are consistent with both scenarios and do not allow an unambiguous conclusion.
\end{references}

\end{document}